\documentclass[11pt,a4paper]{article}
              \usepackage[pdfnewwindow]{hyperref}%
              \title{Quantum Mechanics in Pseudotime}
              \author{\sf A. K. Kapoor
                  \\School of Basic Sciences \\
                   Indian Institute of Technology, Bhubneswar\\
                        Bhubaneswar 751007, India\\
                   {\sf akkapoor@iitbbs.ac.in;\, akkhcu@gmail.com}}
\usepackage{amsmath}
\newcommand{\ket}[1]{|#1\rangle}
\newcommand{\bra}[1]{\langle#1|}

\newcommand{\innerproduct}[2]{\langle#1\vert#2\rangle}

\newcommand{\pp}[2][]{\frac{\partial#1}{\partial #2}}

\newcommand{\dd}[2][]{\frac{d#1}{d #2}}

\newcommand{\KET}[1]{\protect{\vert#1\hspace{-1pt}\succ}}
\newcommand{\BRA}[1]{\protect{\prec\hspace{-1pt}#1\vert}}
\newcommand{\INNPRO}[2]{\prec\hspace{-1pt}#1\vert#2\hspace{-1pt}\succ}

\usepackage{mathrsfs}
\begin{document}

\maketitle
\baselineskip=24pt
\vfill
\begin{abstract}
Based on some results on reparmetrisation of time in Hamiltonian path
integral formalism, a pseudo time formulation of operator formalism of quantum
mechanics is presented. Relation of reparametrisation of time in quantum with
super symmetric  quantum mechanics is established. We show how some important
concepts such as shape invariance and tools like isospcetral deformation appear
in pseudo time quantum mechanics.
\end{abstract}
\vfill
\newpage
\section{Introduction}
The classical trajectory of a point particle can be given in terms a
parameter, other than time, to be called pseudotime in this paper. A Lagrangian
formulation, called homogeneous formalism \cite{mercy} is
available and leads to a singular Lagrangian. A Hamiltonian formulation can be
written down using Dirac's treatment of such systems. Quantisation of such a
system can then proceed by setting the fundamental commutator brackets
equal to \(i\hbar\) times the Dirac brackets. A relativistic particle, string
theory, general relativity constitute important examples which are manifestly
invariant under a reparametrisation of time\cite{RepIn}. Many investigations of
reparametrisation in quantum mechanics start with the homogeneous formalism of
classical mechanics and use Dirac quantisation.

Non relativistic point particle is not mainfestly invariant under a
reparametrisation of time. However it found important applications in the path
integral formulation of quantum mechanics\cite{lsch}. Duru and Kleinert
\cite{DK79}
formulated reparamterisation of time within the Hamiltonian path integral
approach and used it as a tool,  along with K-S transformation\cite{KS}, to
obtain a
path integral solution for H atom problem. Following this idea several
authors used Lagrangian as well as Hamiltonian path integral to write path
integral solution for several exactly solvable potential models.

In \cite{akk1} a scheme of setting up Hamiltonian path integral (HPI)
was proposed. Later quantisation scheme in arbitrary coordinates using
reparametrised Hamiltonian path integral (RHPI) and it was found that
correct quantisation could be carried out without a need  to add
\(O(\hbar^2)\) terms \cite{akk2}. Further results on RHPI and applications have
been given  in \cite{akk3}.

In this paper we wish to formulate reparametrisation of time in the operator
formalism of quantum mechanics closely following the ideas of
reparametrisation in the path integral approach. The central results, that we
need from earlier works, are Duru Kleinert formula and operator ordering. These
are summarised in  Sec.\ref{DKHPISec}.

Path integral representation derives its importance from the fact
that is it can be used as scheme of quantisation too. It is this fact
that will be of crucial importance to us in this paper.

In this article we are concerned with reparametrisation in operator
formalism of nonrelativistic quantum mechanics.
We will connect work on Hamiltonian path integral with some well known results
in quantum mechanics of one dimensional systems. It should then be possible to
extend those results to higher dimensional systems.

The aim of this paper is to  propose a pseuto-time formalism of quantum
mechanics and to link available results of Hamiltonian path integrals to
corresponding results in  the operator formalism and is organised as follows.
The homogeneous
formulation of classical mechanics of a point particle is briefly described in
the next section. Section 3 summarises results about using reparametrisation
of time  in the Hamiltonian path integral approach. In next section several
know results is established with pseudotime quantum mechanics.
The reparamterisation of time appears to be intimately related to supersymmetric
 qunatum mechanics (SUSYQM)\cite{Khare}. We to establish a connection of shape
invariance, isospcetral deformation \cite{Isospec} with pseudo-time formalism.
The quantum Hamilton Jacobi equation(QHJ)\cite{QHJ}, \cite{QHJus} is related to
pseudo time quantum mechanics in a simple and straight forward manner.
A connection  with exceptional polynomials \cite{Camran} is briefly mentioned.
In Sec. 5 the mathematical structure of quantum theory  in pseudo-time is
presented.
Several possible different  interpretations of the mathematical formulation of
pseudo-time quantum mechanics are outlined in  Sec.6. Directions for possible
further study and concluding remarks are given the last section of this paper.

\section{Reparametrisation in classical mechanics}
In this section we briefly describe the homogeneous formalism\cite{mercy}.
In classical mechanics the trajectory of a point particle is found by solving
the Hamilton's equations for generalised coordinates and momenta as functions
of time.
\begin{equation}\label{EQ01}
    q= q(t), \quad p=p(t), \qquad t_1 \le t \le t_2.
\end{equation}
These are parametrised curves in phase space. The trajectories can also be
specified by choosing another parameter \(\sigma\), to be called  pseudo-time.
In this description the coordinates and momenta are to be solved as functions
of \(\sigma\) and the answer is to be written as
\begin{equation}\label{EQ02}
    q= q(\sigma), \quad p=p(\sigma), \qquad \sigma_1 \le \sigma \le \sigma_2.
\end{equation}
To connect with \eqref{EQ01}, we require another equation of the form
\begin{equation}\label{EQ03}
     t= t(\sigma), \qquad t(\sigma_1)=t_1, t(\sigma_2) = t_2.
\end{equation}
treating \(t\) like a coordinate. For this purpose, one adds an equation for
evolution of \(t\) in pseudotime
\begin{equation}\label{EQ04}
    \dd[t]{\sigma} = f(q(t))
\end{equation}
where the function \(f\) is a positive function of coordinates \(q\) and will
be called local time scaling function (LTSF).
A Hamiltonian formalism can be set up by using the pseudo Hamiltonian
\begin{equation}\label{EQ05}
  {\mathscr H }_E = f(q) (H-E)
\end{equation}
The pseudo energy is a constant of motion for dynamics in pseudo time
\(\sigma\). The Hamilton's equations in pseudo time, using pseudo
Hamiltonian\eqref{EQ05},  give the dynamics correctly when the pseudo energy
is set equal to zero. It may be noted that setting pseudo energy equal to zero
means using the energy conservation equation $H(q,p) -E = 0$.
Thus classical equations of motion can be set up equally well in pseudotime
without changing physical content.

\section{Reparametrisation in path integral formalism\label{DKHPISec}}
         In 1979 Duru and Kleinert used K-S transformation and Hamiltonian path
         integral representation in pseudo time to arrive at an exact path
         integral solution for  H-atom problem \cite{DK79}. Crucial to their
         derivation was
         a formula which we call Duru Kleinert formula, see \eqref{DK}.
         This formula related
         the path integral in time \(t\) to Hamiltonian path integral in
         pseudo time, and was derived by Duru and  Kleinert by means of formal
         manipulations. It was therefore a fortunate circumstance that
         it worked for H atom.

         Later the same approach was employed and relation between
         path integrals in time \(t\) and pseudo
         time \(\sigma\) was carefully derived using the accepted rules for
         path integrals within the time slicing  approach\cite{hoino}. In
         general, terms of order \(\hbar^2\) were required to be added
         to the potential when writing setting up  Duru-Kleinert formula.
         A number of authors used Hamiltonian, as well as Lagrangian,
         path integral formulations to
         obtain exact solution for the propagator for many problems \cite{more}.
         It may be recalled that previous to the work by Duru and
         Kleinert, only a small class of quantum mechanical problems could
         be solved using the path integral representation \cite{early}. Use of
         reparametrisation opened the way to obtain  path integral
         solution for large class of problems known to be
         exactly solvable.

         It may be mentioned that reparametrisation, articles on the path
         integral approach used a
         few other techniques, notably addition of new degrees of
         freedom, for arriving at the solutions of potential problems in
         quantum  mechanics.

         It has been known that a careful point transformation, and also use of
         Hamiltonian path integral to quantise  a system in
         non-Cartesian coordinates require addition of \(O(\hbar^2)\)
         terms to  the classical Hamiltonian. It is to be noted that the same
         problem also appears when we attempt to use canonical
         quantisation directly in the
         non-Cartesian coordinates \cite{akk1}. In a series of papers on
         Hamiltonian path integral quantisation, it was shown that
        \(O(\hbar^2)\) terms were
         not required if one suitably that combined reparametrisation with
         the Hamiltonian path integral. Thus reparamterisation of time seems to
         be intimately tied to point transformations in Hamiltonian path
         integrals.

        We recall a few central points and quote a few results from a
        previous study of  Hamiltonian path integral quantisation within
        time slicing approach.

\paragraph*{Hamiltonian path integral\\}
Given a classical Hamiltonian \(H(q,p)\), a scheme of constructing a
particular path integral has been suggested and investigated  in \cite{akk1}.
Using a particular form, \((qt\!\parallel\!q_0t_0)\), for  short time
propagator, Hamiltonian  path integral (HPI) \( K_H(qt,q_0t_0)\) was defined as
summation over all paths from \(q_0\) at time \(t_0\) to \(q\) at time \(t\).
However this did not lead to correct Schrodinger equation in non Cartesian
coordinates.

In \cite{akk2} another path integral representation was introduced by
means for Duru-Kleinert formula
\begin{equation}
 {\mathscr K(qt, q_00)} \stackrel{\rm def}{=}
\int_{-\infty}^{+\infty} \left(\frac{dE}{2\pi\hbar} \right)
\exp(-iEt/\hbar)\int_0^\infty [\sqrt{f(q)f(q_0)}] \times
K_{{\mathscr H}_E}(q\sigma,q_00) d\sigma. \label{DK}
\end{equation}
Here  the expression\(K_{{\mathscr H}_E}(q\sigma, q_0\sigma_0)\), appearing in
the right hand side is the HPI constructed using the pseudo Hamiltoinian
\({\mathscr H}_E= f(q) (H-E)\) corresponding to the
LTSF function \(f(q)\) and will be called reparametrised Hamiltonian path
integral (RHPI). It may be noted that
RHPI reduces to HPI for the special case of LTSF \(f(q)=1\).

The \({\mathscr K(qt, q_00)}\) defined by \eqref{DK} will be called DK
propagator and, to simplify the notation,  its dependence on Hamiltonian
\(H(q,p)\) and LTSF \(f(q)\) will not be shown explicitly .

We will now give a brief summary of main results obtained earlier.

\paragraph*{Operator ordering\\}
Here we list the operator ordering implicit in the above discussion for
the special case of potential problems in \(n\)- dimension.
Corresponding statements for  more general case can be written down but
will not be required here. The Hamiltonian function \(H({\bf x,p})\)
\footnote{Here {\bf x} denotes Cartesian coordinates} will be
assumed to be of the form.
\begin{equation}
  H({\bf x,p}) = \frac{{\bf p}^2}{2m} + V({\bf x})
\end{equation}
As already remarked the HPI obeys the correct Schr\"{o}dinger  equation and
the corresponding operator is
\begin{equation}
  \widehat{H} = -\frac{\hbar^2}{2m} \nabla^2 + V({\bf x}).
\end{equation}

The RHPI \(K_{{\mathscr H}_E}\), satisfies the Schr\"{o}dinger equation with
pseduo Hamiltonian operator
\(\widehat{\mathscr H}_E\) given by
\begin{equation}
  \widehat{\mathscr H}_E = \frac{1}{2m}{\bf \nabla}f({\bf x}){\bf \nabla} +
f({\bf x})\Big( V({\bf x})-E)\Big).
\end{equation}
\paragraph*{Normalisation of RHPI \(K_{{\mathscr
H}_E}\):}
In \cite{akk3} is was shown that the path integral RHPI \(K_{{\mathscr
H}_E}\) appearing in the right hand side of \eqref{DK} satisfies the
Schr\"{o}dinger equation
\begin{equation}
  i\hbar \dd{t}K_{{\mathscr H}_E}({\bf x}t, {\bf x}_0t_0) = \widehat{H}_E
  K_{{\mathscr H}_E}({\bf x}t,{\bf x}_0t_0)
\end{equation}
and has the normalisation
\begin{equation}
\lim_{t\to t_0} K_{{\mathscr H}_E}({\bf x}t,{\bf x}_0t_0)= (f({\bf x}))^{-1}
\delta^{(n)}({\bf x}-{\bf x}_0).
\end{equation}
Thus it is seen that out that the path integral RHPI constructed
with pseudo Hamiltonian is has a normalisation different from what
is required. This explains appearance factor \(\sqrt{f({\bf x})f({\bf x}_0)}\)
in the right hand side of Duru-Kleinert formula.

\paragraph{DK Propagator:}

\begin{equation}
  \widetilde{H}_E = \frac{1}{\sqrt{f}}\left(\frac{1}{2m}{\bf \nabla}f({\bf
x}){\bf \nabla} + f({\bf x})\Big(V({\bf x})-E\Big) \right)\frac{1}{\sqrt{f}}
\end{equation}
Writing this last expression  as \(\widetilde{H}_E\equiv\widetilde{H} -E\), the
expression for
\(\widetilde{H} -E\) can be rearranged in alternate forms
\begin{eqnarray}
  \widetilde{H}
  &=& \frac{1}{f}\frac{1}{2m}(\nabla -{\mathbf w})f({\bf
x})(\nabla-{\mathbf w}) + V({\bf x})\\
  &=& \frac{1}{2m}(\nabla + {\bf w})(\bf \nabla-{\bf w}) +
       V({\bf x})\\
  &=& \frac{1}{2m}\nabla^2 + V(x) + \Delta V
\end{eqnarray}
where \(w_k = \partial_k \Omega({\bf x}), \quad \Omega(x) =\frac{1}{2}\ln f(
{\bf x})\) and
\begin{eqnarray}
  \Delta V = \frac{\hbar^2}{2m} \Big\{(\nabla \Omega(x))^2 -\nabla^2 \Omega(x)
 \Big\}
\end{eqnarray}
The DK propagator \({\mathscr K({\bf x}t,{\bf x}_0t_0})\) of \eqref{DK} obeys
the
Schr\"{o}dinger equation with $\widetilde{H}$ as the Hamiltonian
\begin{equation}
  i\hbar\dd[{\mathscr K({\bf x}t,{\bf x}_0t_0)}]{t}
  =  \widetilde{H}  {\mathscr K({\bf x}t,{\bf x}_0t_0)}
\end{equation}
and is normalised as
\begin{equation}
  {\mathscr K({\bf x}t,{\bf x}_0t_0)}|_{t=t_0} = \delta({\bf x}-{\bf x}_0).
\end{equation}

\paragraph*{Qunatisation in arbitrary coordinates:\\}
Identifying HPI \(K_H\) as propagator for the corresponding quantum
problem leads to correct quantisation scheme in Cartesian coordinates.
However the same scheme, when used for quantisation in non-Cartesian
coordinates, did not give correct quantisation  scheme. It was became necessary
to add \(O(\hbar^2)\) terms to the Hamiltonian\cite{akk1}.

This problem of appearance of \(O(\hbar^2)\) terms
is not specific to the scheme that was   used in reference \cite{akk1}.
A need for these   extra \(O(\hbar^2)\) terms has been  well known   in
the Hamiltonian path   integral   literature. In fact the problem
reappears even in canonical qunatisation scheme. In a general case
this difficulty of canonical qunatisation is masked by ordering
problems  and can be seen most clearly in polar coordinates in two
dimensions, a model not having any ordering problem for the
Hamiltonian.\cite{akk1, lsch}.

{\it It was demonstrated in reference \cite{akk2}, that the
DK-propagator with the choice,  \(f(q)= \rho(q)\) as LTSF function, leads to
correct quantisation in arbitrary coordinates without need to add any
\(O(\hbar^2)\) terms.} This  scheme worked with the {\it classical Hamiltonian}
\(H(q,p)\)
directly in arbitrary coordinates; setting up a Hamiltonian path integral in
Cartesian coordinates and changing variables was not required.

\section{Connection of RHPI with operator formalism}
In this section we establish relationship  of some of the important concepts in
operator formalism and DK propagator.
\paragraph*{Supersymmetric quantum mechanics\\}
Supersymmetric quantum mechanics  has been an active area of research for
several decades and the concept of shape invariance continues to attract a great
deal of attention.
It is easy to see the place that supersymmetric partners and shape invariance
have in the pseudotime path integral framework.
Starting from a free particle Hamiltonian \(V(x)=0\) to set up a
DK propagator, scaling functions \(f(x)\)
and \(1/f(x)\) lead to quantum system whose time evolution is governed by
Hamiltonian $H_+$ and $H_-$, respectively which are given by
\begin{equation}\label{DT1}
  H_\pm = \frac{p^2}{2m} + \frac{\hbar^2}{2m}(w^2 \mp w^\prime),
\end{equation}
where  $w=\dd[\Omega]{x}$ and $\Omega = \ln f$.
\begin{eqnarray}\label{DT2}
  H(q,p) \stackrel{f}{\longrightarrow} \text{DK Propagator for } H_+ \\
  H(q,p) \stackrel{1/f}{\longrightarrow} \text{DK Propagator for }
H_-\label{DTB}
\end{eqnarray}

In the terminology of SUSYQM, the above two Hamiltonians will be recognised as
supersymmetric partners with \(w(x)\) playing the role of superpotential.

\paragraph*{Darboux transformation}
The Darboux transformation \cite{darboux} gives a relation between the
eigenfunctions of SUSY
partner Hamiltonians \(H_\pm\). The propagators for \(H_\pm \) are both related
to free particle HPI in pseudo time with scaling functions \(f(x)\) and
\(1/f(x)\) respectively, see \eqref{DT1}-\eqref{DT2}.

The Darboux transformation and its generalisations by Crum and by Krien
are powerful results that have found large number of applications to several
areas including exactly integrable models.  Darboux's result  is implicitly
contained in Duru Kleinert formula. It is of interest to establish at a direct
and explicit correspondence between Darboux transformation and DK formula.

\paragraph*{Isospectral deformation of a potential\\}
The isospectral deformation of a potential  \(V(x)\) generates a new potential
having exactly same spectrum as the original potential \(V(x)\). This process
makes use of results from SUSYQM.

Consider a model with potential \(V(x)\) corresponding to superpotential
\(w(x)\). We now consider a two RHPI with $H(q,p)$ as free particle with
certain scaling functions \(f(x)\) and \(1/f(x)\). This will result in
DK-propagator for potentials \(V_\pm(x)\).  Demanding that \(V_+(x)\) coincide
with \(V(x)\), a solution for \(f(x)\) will lead to \(V_-(x)\) which will be the
isospectral deformation of the original potential
\(V(x)\). The steps for arriving at the required solution for \(f(x)\) will
closely follow the steps known in the literature for the isospectral
deformation and no further explanation is required.
\paragraph*{Exceptional Polynomials}
In an an earlier paper \cite{akk4eop}, a systematic procedure for deformation
of radial oscillator potential was given in the framework of QHJ was
presented. It was found  that demanding  shape invariance be preserved under
the deformation, led to the  isospectral shift of the radial oscillator
potential. All these steps can in principle be translated and followed in the
pseudotime formalism as presented here. Starting directly form the
differential equation for classical orthogonal polynomials and using using
pseudo time framework,  to will be interesting to find a direct route to the
exceptional polynomials.

\paragraph*{Time dependent sypersymmetry}
Time dependent supersymmetry and time dependent Darboux transformation have
been studied\cite{Bagrov}. These studies  will be connected with
reparametrisation of time
with LTSF which is a function of time, \(\dd[t]{\sigma}=f(q,t)\). The basic
equations in the Hamiltonian path integral formalism  will then  have
correspondence with equations of \cite{Bagrov}.

\paragraph*{Quantum Hamilton Jacobi equation\\}
Quantum Hamilton Jacobi formalism provides a scheme of computing energy
eigenvalues without solving for wave functions \cite{QHJ} and has been
studied extensively \cite{QHJus}
If we substitute \( \psi(x) = \exp(iS(x)/\hbar)\) the
Schr\"{o}dinger equation for a potential problem \(V(\bf x)\)
\begin{equation}
  -\frac{\hbar^2}{2m}\nabla^2\psi({\bf x}) - V(x) \psi({\bf x}) = E \psi({\bf
        x}).
\end{equation}
gets transformed  into an equation for \(S({\bf x})\)
\begin{equation}
  \frac{1}{2m} (\nabla S({\bf x})\big) ^2 +  \frac{i\hbar}{2m} \nabla^2 S({\bf
x}) + V({\bf x}) - E =0. \label{QHJ}
\end{equation}
This equation is known as the quantum Hamilton Jacobi (QHJ) equation for
potential \(V({\bf x})\). Consider the Schrodinger equation
  \begin{equation}
  -\frac{\hbar^2}{2m}\nabla^2 \psi({\bf x}) + V(x) \psi({\bf x})
= E \psi({\bf x}).
\end{equation}
If we set up DK propagator  and demand that  LTSF \(f({\bf x})\)
be such that \(\psi({\bf x})= \) constant is a solution of Schrodinger
equation for \(\widetilde{H}\), we get the following equation for the scaling
function \(f({\bf x})\)
\begin{equation}
 \frac{\hbar^2}{2m}\big[ (\nabla \Omega(x))^2 - \nabla^2 \Omega (x) \big] +
V({\bf x}) - E =0.
\end{equation}
Identifying \(\Omega({\bf x})\) with \(i S({\bf x})/\hbar\), the above equation
becomes identical with the QHJ equation \eqref{QHJ}.

\section{General structure of quantum mechanics in pseudotime}


In this section we will work with non-Cartesian coordinates coordinates and
canonical momenta \({\bf q,p}\) in place of Cartesian coordinates.
The symbol \(\rho({\bf q})\) will denote the volume element  defined by \(d{\bf
x} = \rho({\bf q})d{\bf q}\). The states of system will be elements of
Hilbert space \(\mathcal H\) of all square integrable functions \(f({\bf q})\):
\[\int |\innerproduct{q}{\psi}| \rho(q)\, dq  < \infty \]

Given a system with classical dynamics governed by Hamiltonian
function
  \(H({\bf q,p})\) and a scaling function \(f({\bf q})\), introduce pseudotime
   \(\sigma\) and pseudo Hamiltonian \({\mathscr H}_E\)
\begin{eqnarray}
 \dd[t]{\sigma} &=& f({\bf q}),\\
    {\mathscr H}_E &=& f({\bf q})(H({\bf q,p})-E)=
f({\bf q})({\mathscr H}- E)\label{psHaml},\\
    \text{ where }   {\mathscr H} &=& f({\bf q}) H({\bf q,p}).
\end{eqnarray}

We will work in the Heisenberg picture. Therefore the propagator
  will be given by \(\innerproduct{qt}{q_0t_0}\), where \(\ket{qt}\) is
  eigenvector of the position operator \(\hat{q}(t)\) at time \(t\):
  \begin{equation}
    \hat{q}_k(t)\ket{{\bf q}t} = q_k\ket{{\bf q}t}, k=1,2,\ldots.
  \end{equation}
  A path integral representation for propagator is constructed out of
  short time propagator \(({\bf q} t\vert {\bf q}t_0)\) in the usual fashion by
  using time slicing approach and summing over all paths.

We recall \eqref{DK} and the comment on normalisation of  RHPI. It has been
noted that the HPI in real time \(t\) and RHPI in pseudo time \(\sigma \) are
normalised differently, and that one needs to multiply by
factor \(\sqrt{f({\bf x}) f({\bf x}_0)}\) at the end. We introduces a second
set of position eignevectors   \( {\mathscr B}=\{\KET{\bf
q}\} \) defined by
\[\KET{\bf q}= (f({\bf q}))^{-1/2}\ket{q}\]
which are normalised differently:
\begin{equation}
  \INNPRO{\bf q}{\bf q_0} =  (f(\bf q))^{-1} \rho^{-1}({\bf q})
    \delta^{(n)}(q-q_0).
\end{equation}
The completeness relations now take the form
\begin{eqnarray}
  \int \rho({\bf q})\ket{\bf q}\bra{\bf q} dq = \hat{I},\qquad
  \int \rho(\bf q)f(\bf q)\KET{\bf q}\BRA{\bf q}= \hat{I}.
\end{eqnarray}
The new eigenvectors of position \(\KET{\bf q}\) span a different Hilbert space
\({\mathcal H}_f\) with changed scalar scalar product
\(\innerproduct{\psi}{\phi}\) defined as:
\begin{equation}
 \int  \psi({\bf q})\phi({\bf q}) \rho({\bf q}) \,d{\bf q}  \longrightarrow
   \int  \psi({\bf q})\phi({\bf q}) \rho({\bf q}) f({\bf q})\, d{\bf q}. \
\end{equation}
Thus we interpret the change to basis \(\KET{\bf q}\) as a switch to
different Hilbert space of  functions square integrable with a new measure
\(\rho(q)f(q)\).

The process of setting up the RHPI with LTSF \(f({\bf q})\) can now be regarded
as consisting of following steps.

Noting that
\(\innerproduct{\bf q}{\psi}\in {\mathcal H}\)
and \(\INNPRO{\bf q}{\psi}\in {\mathcal H}_f\) representing a state vector
\(\ket{\psi}\)  in the two Hilbert spaces are related by
\begin{equation}
  \INNPRO{{\bf q}}{\psi} = (f({\bf q}))^{-1/2} \innerproduct{\bf q}{\psi},
  \label{EQ33}
\end{equation}
we define a mapping of the operators in the two Hilbert spaces by
\begin{equation}\label{EQ34}
 {\mathscr X}_f=  (f({\bf q}))^{-1/2}\, \hat{X}\, (f({\bf q}))^{1/2}.
\end{equation}
The vector space equations will then be preserved when a transition is made from
the original Hilbert space to another one labelled by \(f({\bf q})\).

If \(\hat{X}\) is written as an ordered expression  \(\hat{X}\big({\bf
q},\pp{\bf q}\big)\), the above equation translates into
\begin{equation}
 {\mathscr X}_f= X ({\bf q},{\bf D})
\end{equation}
where
\begin{equation}
D_k =\partial_k  + \omega_k \quad \text{and}\quad \omega_k=\frac{1}{2}\pp[\,(\ln
f)]{q_k}.
\end{equation}

Setting up of the the propagator as RHPI and using DK propagator can now be
interpreted as a sequence of the following steps:
\begin{enumerate}
\item Change from original Hilbert space \({\mathcal H}\)  to a another
      Hilbert space \({\mathcal H}_f\) labelled by the LTSF \(f({\bf q})\), and
      defined by a different square integrability requirement
      \begin{equation}
        \int |\psi({\bf q})|^2 \rho({\bf q})dq < \infty \longrightarrow
        \int |\psi({\bf q})|^2 \rho({\bf q})f({\bf q})d{\bf q} <      \infty.
      \end{equation}

\item Set up the propagator in the new Hilbert space \({\mathcal H}_f\)
      as a path integral RHPI in pseudotime \(\sigma\).
\item Revert back to original Hilbert\({\mathcal H}\) space by using DK
propagator for time  evolution in time \(t\).
\end{enumerate}
\section{Interpretations of pseudotime quantum mechanics}
So far we have looked at the mathematical structure of quantum theory as
suggested by Hamiltonian path integral in pseudo time. Now we discuss
a few different possible interpretations of our equations
in a manner which incorporates reparametrisation of time in quantum theory.

The classical formulations in  different pseudo times are all
equivalent in the sense that they give rise to the same trajectory.
In quantum theory they solutions of dynamical equations look
different for different choices of pseudo times because of appearance of
\(O(\hbar^2)\) terms appearing in the potential. For example a free particle in
real time \(t\) will, in pseudo time, appear as a particle moving in a non
constant potential \(\Delta V\).

Simplest interpretation would of our equations will  be that, for a particular
choice of Hamiltonian \(H({\bf q,p})\), only DK
propagator is to be regarded as having physical significance. The RPHI is
introduced
solely for technical purposes for setting up the propagator and that  it
appears in intermediate steps of  quantisation and evaluation of the
propagator using path integrals. A possibility along these lines is that  for
each choice of coordinates \({\bf q}\), there is one ( or more )  preferred
choice(s) of LTSF \(f({\bf q})\) determined by the requirement \(\Delta V=0\),
as is indicated by quantisation in arbitrary coordinates.

Another way to interpret the equations and \(\Delta V\) term is that the
quantum mechanical equations are not invariant under the reparametrisation of
time. The DK propagator for different LTSF function \(f({\bf q})\) will give
the same results  if we {\it subtract a  \(\Delta V\)
term} from the potential while writing the pseudo Hamiltonian. This would not be
a desirable feature for theories  which have reparametrisation
invariance built in the  classical theory  and the symmetry needs to
be preserved in the quantum theory too.

In a third approach  to interpretation, one can view the quantum theory as
fundamental and interactions being linked in some way to choice of physical
time, or to a particular frame of reference. In this case one would still need
to discover a manifestly covariant form of equations in quantum theory.
\cite{jafar}

\paragraph*{A gauge covariant formalism for pseudo time dynamics:}
There is yet another approach to setting up equations in pseudo time. The
states and operators representing dynamical variables transform according to the
rule in \eqref{EQ33} and \eqref{EQ34}.
In coordinate representation this is like a nonunitary, but invertible, gauge
transformation
\begin{equation}
  \psi(x) \to \psi_f(x)= \exp(\Omega(x)) \psi(x), \label{GT}
\end{equation}
where \(\Omega\) is a real function of \(x\).

Recall that we are free to choose the overall normalisation of the wave function
in any fashion we like, \(\psi(x)\) and \(N\psi(x)\) give rise to same physics.
 The transformation \eqref{GT} amounts to gauging this freedom in choice of
normalisation constant. This, in fact, is close to the original suggestion of
Hermann Weyl and was used by him in a different context and with a different
motivation.\cite{Weyl}

Thus one can think of introducing a gauge field required to maintain
the new gauge invariance. Such a 'potential field' can always be introduced as a
mathematical construction to do book keeping and the potential
having different values in different gauges (pseudo times).
Whether this 'field' can be assigned dynamical properties can only decided by
further investigation and confirmation of its existence by experiments.

\section{Concluding remarks}
It is obvious that SUSYQM has initmate connection with reparametrisaton of time
in quantum mechanics. The tower of SUSY partners correspond to choices
LTSF \(f^n(x)\) for different values of \(n\). It has been shown in this paper
that this connection becomes transparent when one uses the Hamiltonian path
integral formalism of quantum mechanics. Since the path integral approach is
not restricted to one dimension, it offers a possibility of formulating SUSYQM
in higher dimensions. In \cite{akkSI} the concept of shape invariance has
been extended to quantum mechanics in arbitrary dimensions. However, a
generalisation of a few other results such as intertwining of partner
potentials, is needed in order to carry forward work done in SUSYQM one
dimension.
Darboux transformation is powerful tool and has a wide ranging applications in
different areas of mathematical physics and exactly solvable models.
It will be useful to further study connection  of Darboux transformation with
RHPI.

An area where results of our work on reparamterisation will be useful is  to
the systems where the action has singularities \cite{jafar} and standard path
integral needs to be regularised. Working in pseudotime, with a suitable choice
of LTSF, can be used to regularize singularities. Use of path integral to
quantise and shifting to operator formalism should makes it technically
easier as compared to using the path integral formalism alone.

One of the important approaches to reparametrisation is
using the homogeneous formalism and applying Dirac quantisation. The Dirac
canonical quantisation fixes commutators, leaving questions related to ordering
open. The Hamiltonian path integral quantisation approach followed here is
powerful framework and goes beyond canoincal quantisation. It determines an
operator ordering too. Besides fixing  operator ordering, the formalism proposed
here gives a way of relating results of pseudotime quantum mechanics with those
in time \(t\); an explicit formula, DK formula, has  been written down to
relate the time evolutions in time \(t\) and pseudotime. Though the
ordering scheme depends on the way Hamiltonian path integral is set up,
the scheme  of \cite{akk2,akk3} appears to useful for establishing connections
with  SUSYQM and other areas in the literature. Further investigation of
RHPI and it  correspondence with operator formalism appears to be promising for
a study of several areas of exactly solvable systems and
systems with reparametrisation invariance.

\noindent
{\it Acknowledgement:} I thank Pankaj Sharan for fruitful discussions.

\end{document}